# Unified Description of Dark Energy and Dark Matter

Walter Petry
Mathematisches Institut der Universitaet Duesseldorf, D-40225 Duesseldorf
E-mail: wpetry@meduse.de
petryw@uni-duesseldorf.de

Abstract: Dark energy in the universe is assumed to be vacuum energy. The energy-momentum of vacuum is described by a scale-dependent cosmological constant. The equations of motion imply for the density of matter (dust) the sum of the usual matter density (luminous matter) and an additional matter density (dark matter) similar to the dark energy. The scale-dependent cosmological constant is given up to an exponent which is approximated by the experimentally decided density parameters of dark matter and dark energy. This yields that dark matter is one third of dark energy for all times implying an explanation of the coincidence of dark matter and dark energy. In the final state, the universe becomes dark, consisting of dark matter and dark energy.

## 1. Introduction

The WMAP temperature data, the small scale CMB data, the large-scale structure observations and the supernovae distance-redshift relationship give the best fit values for the cosmological parameters (see. e.g. Bludman [1] especially Spergel et al. [2]):

$$\Omega_l = 0.0462 \pm 0.00115, \quad \Omega_d = 0.233 \pm 0.013, \quad \Omega_\Lambda = 0.721 \pm 0.015, \tag{1.1}$$

where $\Omega_l$ denotes the present density parameter of the luminous matter, $\Omega_d$ of the dark matter and $\Omega_\Lambda$ of the dark energy. The spatial curvature is in agreement with a flat space (see. e.g. [1,2]).

The standard cosmological models assume a homogeneous, isotropic universe where the source is radiation, matter (dust) and dark energy. The last one is in general given by the cosmological constant. Luminous and dark matter are described up to a constant by the same formula. Such models are studied by the use of Einstein's general theory of relativity and by the theory of gravitation in flat space-time (see [3]). In the beginning of the universe the two theories yield quite different results. The general theory of relativity starts from a singularity, i.e. the densities of matter and radiation are infinite and then the universe begins to exist. The singularity is called the big bang of the universe. The theory of gravitation in flat space-time starts from a state where no matter and no radiation exist and all the energy is gravitational energy. Then, radiation, matter and dark energy arise and increase in the course of time. The sum of all the energies of matter, radiation, dark energy and gravitational energy is conserved by virtue of the conservation law of the total energy-momentum. At some time, the densities of radiation and matter become very high but they remain finite which corresponds to the big bang of the general theory of relativity. Later on, the cosmological models of the two theories agree with one another if a flat space for Einstein's theory is assumed.

Subsequently, the covariant theory of gravitation in flat space-time is used (see[3]). This gravitational theory is based on a flat space-time metric and gravitation is described by a field, analogously to Maxwell's theory. This theory of gravitation gives the same results as Einstein's general theory of relativity to the experimentally needed accuracy for the following effects: gravitational redshift, light deflection, perihelion precession, radar time delay, post-Newtonian approximation, gravitational radiation, and the precession of the spin axis of a gyroscope in the orbit of a rotating body. As already stated the theory of gravitation in flat space-time gives non-singular, cosmological models and furthermore, Birkhoff's theorem is not valid in contrast to Einstein's general theory of relativity. The space of the universe is flat by flat space-time theory of gravitation which is also suggested by recent observations (see e.g. [1,2]).

In this paper, dark energy of the universe is assumed to be vacuum energy. The energy-momentum tensor of this vacuum energy is given by a cosmological constant which depends on the scale-factor of the universe.The covariant equations of motion imply that the density of matter (dust) is the sum of two different expressions of densities.The first one is the usual expression and it is interpreted as luminous matter. The second expression of the density is related to the vacuum energy density where the function of the scale-dependent cosmological constant contains a constant of integration. This constant is approximated by the use of the density parameters of dark matter and dark energy.The second expression of the matter density is interpreted as dark energy. It is given up to a constant by the same formula as dark energy. This result gives an explanation of the coincidence problem of dark matter and dark energy. A homogeneous, isotropic universe with these densities of radiation, luminous matter, dark matter, and dark energy is studied. The beginning of the non-singular universe can be described analogously as by the standard cosmological model where radiation, luminous matter, dark matter, and dark energy arise. In the course of time the densities of radiation and luminous matter become very high but they remain finite. Later on, dark matter and dark energy predominate radiation and luminous matter and an analytic solution for this case is given. In the final state, the universe is dark and it consists of dark matter and dark energy.

## 2. Cosmological Model

Let us start from the theory of gravitation in flat space-time which has been studied in several papers (see e.g. [3-11]). Flat space-time theory of gravitation uses a flat space-time background metric

$$(ds)^2 = -\eta_{ij} dx^i dx^j. \tag{2.1}$$

Gravitation is described by gravitational potentials $g_{ij}$ satisfying covariant (with regard to (2.1)) field equations of the form

$$\widetilde{R}^i{}_j - \frac{1}{2}\delta^i{}_j \widetilde{R}^k{}_k = 4\kappa T^i{}_j \tag{2.2}$$

where $\widetilde{R}^i{}_j$ is a covariant differential operator of order two, $T^i{}_j$ denotes the total energy-momentum tensor inclusive that of the gravitational field and $\kappa = 4\pi k/c^4$ ($k$: gravitational constant). In addition to the field equations (2.2), the conservation law of the total energy-momentum holds, i.e.

$$T^k{}_{i/k} = 0 \tag{2.3}$$

where the bar "/" denotes the covariant derivative. The field equations (2.2) and the conservation law (2.3) give the covariant equations of motion for the total matter, i.e.

$$\overset{m}{T}{}^j{}_{i/j} = \frac{1}{2} g_{jk/i} \overset{m}{T}{}^{jk} \tag{2.4a}$$

where

$$\overset{m}{T}{}^{ij} = g^{jk} \overset{m}{T}{}^i{}_k \tag{2.4b}$$

is the symmetric matter tensor. The inversion also holds, i.e. the field equations (2.2) and the equations of motion (2.4) imply the conservation law (2.3) of the total energy-momentum tensor. Furthermore, the proper time $\tau$ is defined by

$$c^2 (d\tau)^2 = -g_{ij} dx^i dx^j. \tag{2.5}$$

These results can be found e.g. in the papers [4,6] and the cited literature therein. Define

$$\eta = \det(\eta_{ij}) \tag{2.6}$$

then the equations of motion (2.4) can be rewritten in the non-covariant form

$$\frac{1}{(-\eta)^{1/2}} \frac{\partial}{\partial x^j} \left( (-\eta)^{1/2} \overset{m}{T}{}^j{}_i \right) = \frac{1}{2} \frac{\partial g_{jk}}{\partial x^i} \overset{m}{T}{}^{jk}. \tag{2.7}$$

The energy-momentum tensor of a perfect fluid is given by

$$\overset{F}{T}{}^i{}_j = (\rho + p) g_{jk} \frac{dx^i}{d\tau} \frac{dx^k}{d\tau} + \delta^i{}_j p c^2 \tag{2.8}$$

and the vacuum energy-momentum tensor has the form

$$\overset{\Lambda}{T}{}^i{}_j = -\frac{\Lambda}{2\kappa} \left( \frac{-G}{-\eta} \right)^{1/2} \delta^i{}_j. \tag{2.9}$$

Here, $\rho$ and $p$ are density and pressure, $\Lambda$ is assumed to be not constant and

$$G = \det(g_{ij}). \tag{2.10}$$

In the case where $\Lambda$ is constant, then (2.9) corresponds to the cosmological constant introduced by Einstein.
The theory of gravitation in flat space–time is applied to homogeneous, isotropic cosmological models (see e.g. [10,11]). Let $x^i$ $(i=1,2,3)$ be the Cartesian coordinates and $x^4 = ct$ and the background metric is given by

$$(\eta_{ij}) = diag(1,1,1,-1). \tag{2.11}$$

The density and pressure are

$$\rho = \rho_r + \rho_m, \quad p = p_r + p_m \tag{2.12}$$

where $r$ and $m$ denote radiation and matter (dust) with

$$p_r = \rho_r/3, \quad p_m = 0. \tag{2.13}$$

Furthermore, let us assume the velocity

$$\frac{dx^i}{d\tau} = 0 \quad (i=1,2,3). \tag{2.14}$$

Then, the gravitational potentials have only diagonal elements, i.e.
$$(g_{ij}) = diag(a^2(t), a^2(t), a^2(t), -1/h(t)) \tag{2.15}$$

It follows
$$(g^{ij}) = diag(1/a^2(t), 1/a^2(t), 1/a^2(t), -h(t)) \tag{2.16}$$

and
$$(-G)^{1/2} = a^3(t)/h^{1/2}(t), \quad (-\eta)^{1/2} = 1. \tag{2.17}$$

The equation (2.5) gives with (2.14) and (2.15)
$$\frac{dx^4}{d\tau} = ch^{1/2}. \tag{2.18}$$

The relations (2.8) and (2.9) imply by the use of (2.14), (2.18), (2.15), (2.12), (2.13) and (2.17) the energy-momentum of the total matter, i.e. perfect fluid (dust and radiaion) and vacuum

$$\overset{m}{T}{}^i{}_j = \overset{F}{T}{}^i{}_j + \overset{\Lambda}{T}{}^i{}_j = p_r c^2 - \frac{\Lambda(a)}{2\kappa} a^3/h^{1/2}, \quad i = j = 1,2,3$$

$$= (\rho_r + \rho_m)c^2 - \frac{\Lambda(a)}{2\kappa} a^3/h^{1/2}, \quad i = j = 4$$

$$= 0, \quad i \neq j \tag{2.19a}$$

Here, it is assumed that $\Lambda = \Lambda(a)$. The energy-momentum tensor of the gravitational field has the form

$$\overset{G}{T}{}^i{}_j = \frac{1}{16\kappa} L_G, \quad i = j = 1,2,3$$

$$= -\frac{1}{16\kappa} L_G, \quad i = j = 4$$

$$= 0, \quad i \neq j \tag{2.19b}$$

where
$$L_G = \frac{1}{c^2} a^3 h^{1/2} \left( -6\left(\frac{\dot{a}}{a}\right)^2 + 6\frac{\dot{a}}{a}\frac{\dot{h}}{h} + \frac{1}{2}\left(\frac{\dot{h}}{h}\right)^2 \right) \tag{2.19c}$$

and the dot "•" denotes the $t$-derivative. The total energy-momentum tensor has the form
$$T^i{}_j = \overset{m}{T}{}^i{}_j + \overset{G}{T}{}^i{}_j. \tag{2.20}$$

The field equations (2.2) imply two nonlinear coupled differential equations of order two of the form

$$\frac{d}{dt}\left(a^3 h^{1/2}\frac{\dot{a}}{a}\right) = 2\kappa c^4\left(\frac{1}{2}\rho_m + \frac{1}{3}\rho_r + \frac{\Lambda(a)}{2\kappa c^2} a^3/h^{1/2}\right)$$

$$\frac{d}{dt}\left(a^3 h^{1/2}\frac{\dot{h}}{h}\right) = 4\kappa c^4\left(\frac{1}{2}\rho_m + \rho_r + \frac{1}{2\kappa c^2} L_G - \frac{\Lambda(a)}{2\kappa c^2} a^3/h^{1/2}\right) \tag{2.21}$$

where (2.13) is used. Let us choose the present time $t_0 = 0$ as initial conditions:
$$a(0) = h(0) = 1, \quad \dot{a}(0) = H_0, \quad \dot{h}(0) = \dot{h}_0. \tag{2.22}$$

Here, $H_0$ is the Hubble constant and $\dot{h}_0$ is a further unknown constant. The conservation law of the total energy (2.3) has by the use of (2.19) and (2.20) the form
$$(\rho_m + \rho_r)c^2 + \frac{1}{16\kappa} L_G + \frac{\Lambda(a)}{2\kappa} a^3/h^{1/2} = \lambda c^2 \tag{2.23}$$

where $\lambda$ is a constant of integration. The equations of motion (2.7) give for $i = 4$ by the use of (2.19a), (2.15), (2.16) and some elementary calculations the relation

$$\frac{d}{dt}\left(\rho_m h^{1/2}\right)+\frac{1}{a}\frac{d}{dt}\left(\rho_r a h^{1/2}\right)+a^3\frac{d}{dt}\left(\frac{\Lambda(a)}{2\kappa c^2}\right)=0. \tag{2.24}$$

In the special case where every expression of the sum vanishes, the standard cosmological model is received with a fixed cosmological constant $\Lambda_0$. The result of this model can be found in many papers ( see e.g. [10,11]). Let us now assume that

$$\frac{d}{dt}\left(\rho_r a h^{1/2}\right)=0 \tag{2.25}$$

then equation (2.25) gives the density of radiation

$$\rho_r = \rho_{r0}/\left(ah^{1/2}\right) \tag{2.26}$$

where $\rho_{r0}$ denotes the present density of radiation. The expression (2.26) is identical with the radiation density of the standard cosmological model. Relation (2.24) has by the use of (2.25) the form

$$\frac{d}{dt}\left(\rho_m h^{1/2}\right)+a^3\dot{a}\frac{d}{da}\left(\frac{\Lambda(a)}{2\kappa c^2}\right)=0. \tag{2.27}$$

Let us assume

$$\Lambda(a)=\Lambda_0 a^{k+1} \tag{2.28}$$

with a cosmological constant $\Lambda_0$ then relation (2.27) is rewritten in the form

$$\frac{d}{dt}\left(\rho_m h^{1/2}\right)+\frac{\Lambda_0}{2\kappa c^2}(k+1)a^{k+3}\dot{a}=0.$$

This differential equation has the solution

$$\rho_m h^{1/2}=C_0-\frac{\Lambda_0}{2\kappa c^2}\frac{k+1}{k+4}a^{k+4}$$

where $C_0$ is a constant of integration. Hence, the matter density (dust) is given with $C_0=\rho_{l0}$ by

$$\rho_m = \rho_{l0}/h^{1/2}-\frac{\Lambda_0}{2\kappa c^2}\frac{k+1}{k+4}a^{k+4}/h^{1/2}. \tag{2.29}$$

The first expression on the right hand side of relation (2.29) is interpreted as luminous matter whereas the second expression gives the dark matter. We must have

$$-4 < k < -1$$

to get positive dark matter density. Hence, the density of dark matter is given by

$$\rho_d = -\frac{k+1}{k+4}\frac{\Lambda_0}{2\kappa c^2}a^{k+4}/h^{1/2}. \tag{2.30}$$

The density of the vacuum energy $\rho_\Lambda$ follows from (2.23) by the use of (2.28)

$$\rho_\Lambda = \frac{\Lambda_0}{2\kappa c^2}a^{k+4}/h^{1/2}. \tag{2.31}$$

The relations (2.30) and (2.31) imply

$$\rho_d = -\frac{k+1}{k+4}\rho_\Lambda$$

for all times. The density parameters (1.1) at present for dark matter and dark energy give the approximation

$$k \approx -1.733 \approx -7/4.$$

Hence, let us assume

$$k = -7/4 \tag{2.32}$$

because the densities of dark matter and dark energy must satify a simple law and the values of (1.1) give only approximations of the density parameters. The value (2.32) implies

$$\rho_d = \frac{1}{3}\rho_\Lambda \tag{2.33}$$

for all times. This relation yields a unification of dark matter and dark energy. Several authors have supposed such a result.

Subsequently, we use the results of the papers [5,11] to study the resulting new cosmological model.
Let us define as usually the density parameters

$$\Omega_r = \frac{8\pi k \rho_{r0}}{3H_0^2}, \quad \Omega_m = \frac{8\pi k \rho_m(0)}{3H_0^2}, \quad \Omega_\Lambda = \frac{\Lambda_0 c^2}{3H_0^2}, \quad \Omega_l = \frac{8\pi k \rho_{l0}}{3H_0^2}, \quad \Omega_d = \frac{1}{3}\Omega_\Lambda \qquad (2.34)$$

where $r, l, d, m$ and $\Lambda$ denote radiation, luminous matter, dark matter, matter (dust) of luminous and dark matter and dark energy. Furthermore, let us define the constants

$$k_0 = \Omega_r + \Omega_l + \Omega_d + \Omega_\Lambda - 1 \qquad (2.35a)$$

and

$$\varphi_0 = 3H_0\left(1 + \frac{1}{6}\frac{\dot{h}}{h}\right). \qquad (2.35b)$$

Then, equation (2.23) implies at present time with the aid of (2.34):

$$\frac{2\kappa c^4 \lambda}{H_0^2} - \left(\frac{1}{2}\frac{\varphi_0}{H_0}\right)^2 = 3k_0. \qquad (2.36)$$

We get by the use of the results of the papers [5,11] with (2.29), (2.30), (2.31) and (2.32) the relation

$$a^3 h^{1/2} = 2\kappa c^4 \lambda t^2 + \varphi_0 t + 1 \qquad (2.37)$$

and the differential equation

$$\left(\frac{\dot{a}}{a}\right)^2 = \frac{H_0^2}{(2\kappa c^4 \lambda t^2 + \varphi_0 t + 1)^2}\left[-k_0 + \Omega_r a^2 + \Omega_l a^3 + (\Omega_d + \Omega_\Lambda)a^{21/4}\right]. \qquad (2.38)$$

To get non-singular cosmological models we must have for all times $t \in R$

$$2\kappa c^4 \lambda t^2 + \varphi_0 t + 1 > 0$$

which is by the use of (2.36) equivalent to

$$k_0 > 0. \qquad (2.39)$$

Therefore, condition (2.39) implies by the use of (2.38) the existence of a smallest positive value $a_1$ of $a(t)$ satisfying

$$k_0 = \Omega_r a_1^2 + \Omega_l a_1^3 + (\Omega_d + \Omega_\Lambda)a_1^{21/4}. \qquad (2.40)$$

Hence, condition (2.39) is necessary and sufficient for the existence of non-singular cosmological models.
The distance of a galaxy with redshift $z$ is given by

$$r = \frac{c}{H_0}\int_a^1 \frac{xdx}{\left(-k_0 + \Omega_r x^2 + \Omega_l x^3 + (\Omega_d + \Omega_\Lambda)x^{21/4}\right)^{1/2}}. \qquad (2.41a)$$

with

$$a = 1/(1+z). \qquad (2.41b)$$

To get a hot universe, we assume as for standard cosmological models

$$k_0 \ll 1. \qquad (2.42)$$

Hence, relation (2.35a) together with (2.39) and (2.42) implies that the sum of all the density parameters is a little bit greater than one. It follows by virtue of $\Omega_r \ll \Omega_l$

$$\Omega_l + \Omega_d + \Omega_\Lambda \approx 1.$$

This relation implies with the aid of the last relation in (2.34) and the approximate value $\Omega_l$ of (1.1)

$$\Omega_\Lambda \approx 0.7154, \quad \Omega_d \approx 0.2385. \qquad (2.43)$$

These parameters are in good agreement with the experimentally received density parameters $\Omega_\Lambda$ and $\Omega_d$ given in (1.1).

For more details of the beginning of the non-singular universe, the results of the papers [5,11] of the standard cosmological models can be considered.

For a sufficiently large function $a(t)$, differences between the standard and the new cosmological model arise. Assume that

$$a(t) \gg \left(\frac{\Omega_l}{\Omega_d + \Omega_\Lambda}\right)^{4/9} \approx 0.26, \quad \text{i.e.} \quad z \ll \left(\frac{\Omega_d + \Omega_\Lambda}{\Omega_l}\right)^{4/9} - 1 \approx 2.85. \tag{2.44}$$

Then, equation (2.38) can be approximated by the use of (2.36) with the restrictions (2.39) and (2.42)

$$\frac{\dot{a}}{a} \approx \frac{H_0}{(\varphi_0 t/2 + 1)^2} (\Omega_d + \Omega_\Lambda)^{1/2} a^{21/8}.$$

The solution of this differential equation with the initial condition $a(0)=1$ is given by

$$a^{21/8}(t) \approx 1 / \left[1 - \frac{21}{8}(\Omega_d + \Omega_\Lambda)^{1/2}\left(1 - \frac{1}{\varphi_0 t/2 + 1}\right) / \left(\frac{1}{2}\frac{\varphi_0}{H_0}\right)\right]. \tag{2.45}$$

It follows from (2.41) under the assumption (2.44) for the distance of a galaxy as function of the redshift:

$$r \approx \frac{8}{5} \frac{c}{H_0} \frac{1}{(\Omega_d + \Omega_\Lambda)^{1/2}} \left((1+z)^{5/8} - 1\right). \tag{2.46}$$

Relation (2.45) requires

$$\frac{21}{8}(\Omega_d + \Omega_\Lambda)^{1/2} \leq \frac{1}{2}\frac{\varphi_0}{H_0}.$$

This inequality yields by (2.35b)

$$\dot{h}_0 \geq 6H_0\left(\frac{7}{4}(\Omega_d + \Omega_\Lambda)^{1/2} - 1\right). \tag{2.47}$$

There are two different cases:

(I) $$\dot{h}_0 > 6H_0\left(\frac{7}{4}(\Omega_d + \Omega_\Lambda)^{1/2} - 1\right). \tag{2.47a}$$

Then, $a(t)$ increases to a finite value

$$a(\infty) = 1 / \left(1 - \frac{21}{8}(\Omega_d + \Omega_\Lambda)^{1/2} / \left(\frac{1}{2}\frac{\varphi_0}{H_0}\right)\right)^{8/21}$$

as $t \to \infty$. This means that the final state of the universe consists only of gravitational energy as in the beginning of the universe, i.e. all the matter (luminous and dark) and dark energy go to zero. Thus, the universe does no more exist.

(II) $$\dot{h}_0 = 6H_0\left(\frac{7}{4}(\Omega_d + \Omega_\Lambda)^{1/2} - 1\right). \tag{2.47b}$$

In this case, we get for sufficiently large $t$

$$a(t) \approx \left(\frac{1}{2}\varphi_0 t + 1\right)^{8/21} \tag{2.48a}$$

and relation (2.37) yields

$$h^{1/2}(t) \approx \left(\frac{1}{2}\varphi_0 t + 1\right)^{6/7}. \tag{2.48b}$$

Hence, dark matter and dark energy go to a finite value whereas luminous matter and radiation go to zero as $t \to \infty$, i.e. the final state of the universe consists of dark matter and dark energy together with gravitational energy. Thus, the universe will exist furthermore. Therefore, case (II) seems to be more natural to describe our universe. In both cases the universe becomes dark.

Summarizing, it follows that dark energy described by vacuum energy with a scale-dependent cosmological constant implies dark matter in addition to luminous matter where dark matter is one third of dark energy. This result solves the experimentally stated coincidence problem not only valid at present but for all times. In the course of time the universe becomes dark. The final state of the universe is dark matter and dark energy whereas luminous matter does no more exist.